\documentclass[twocolumn,showpacs,preprintnumbers,amsmath,amssymb,prb,aps]{revtex4-2} 
\usepackage{epsfig}% Include figure files
\usepackage{epstopdf}
\usepackage{multirow}
\usepackage{graphicx}% Include figure files
\usepackage{dcolumn}% Align table columns on decimal point
\usepackage{bm}% bold math
\usepackage{textcomp}
\usepackage{array}
\usepackage{csquotes}
\usepackage{subcaption}
\usepackage{booktabs}
\usepackage{amsmath}
\usepackage{hyperref}% add hypertext capabilities
\hypersetup{
    colorlinks=true,
    linkcolor=blue,
    filecolor=magenta,      
    urlcolor=red,
    pdftitle={},
    }
    
\begin{document}
\title{\textit{Ab initio} investigation of the lattice dynamics and thermophysical properties of BCC Vanadium and Niobium}
\author{Prakash Pandey$^{1}$}
\altaffiliation{ \url{prakashpandey6215@gmail.com}}
\author{Sudhir K. Pandey$^{2}$}
\altaffiliation{ \url{sudhir@iitmandi.ac.in}}
\affiliation{$^{1}$School of Physical Sciences, Indian Institute of Technology Mandi, Kamand - 175075, India\\$^{2}$School of Mechanical and Materials Engineering, Indian Institute of Technology Mandi, Kamand - 175075, India}

\date{\today}

\begin{abstract}
In the present work, we have performed the phonon dispersion calculations of body-centered cubic Vanadium (V) and Niobium (Nb) with the supercell approach using different supercell size. Using DFT method, the calculated phonon spectra of V and Nb are found to be in a good agreement with the available experimental data. Our calculated results show a \enquote{dip}-like feature in the longitudinal acoustic phonon mode along the $\Gamma$-H high symmetric path for both transition metals in the case of supercell size 4$\times$4$\times$4. However, in supercell size 2$\times$2$\times$2 and 3$\times$3$\times$3, the \enquote{dip}-like feature is not clearly visible. In addition to this, thermodynamical properties are also computed, which compare well with the experimental data. Apart from this, the phonon lifetime due to electron-phonon interactions ($\tau_{\rm{eph}}^{\rm{ph}}$) and phonon-phonon interactions ($\tau_{\rm{ph}}^{\rm{ph}}$) are calculated. The effect of phonon-phonon interactions (PPI) is studied by computing the average phonon lifetime for all acoustic branches. The value of $\tau_{\rm{eph}}^{\rm{ph}}$ of V (Nb) is found to be 23.16 (24.70)$\times 10^{-15}$ s at $100$ K, which gets decreased to 1.51 (1.85)$\times 10^{-15}$ s at 1000 K. The $\tau_{\rm{ph}}^{\rm{ph}}$ of V (Nb) is found to be 8.59 (18.09)$\times 10^{-12}$ and 0.83 (1.76)$\times 10^{-12}$ s at 100 and 1000 K, respectively. Nextly, the lattice thermal conductivity is computed using linearized phonon Boltzmann equation. The present work suggests that studying the variation of phonon dispersion with supercell size is crucial for understanding the phonon properties of solids accurately. 

\end{abstract}

\maketitle

%******************************************************** Introduction ************************************************************
%\setlength{\parindent}{3em}
\section{Introduction}
Phonons play a crucial role in understanding the phononic properties of elements and compounds. The phonon density of states provides the spectrum of energies of phonons, which is essential for calculating the thermodynamic properties of materials. For understanding and calculating the thermodynamic properties of solids, such as free energy, entropy and heat capacity, harmonic phonons provide a valuable framework. Harmonic phonons are non-interacting elementary excitations characterised by their frequencies and have an infinite lifetime\cite{ashcroft}. At the level of the harmonic approximation, individual phonons are considered to be independent quantum mechanical entities. A wide range of extremely important novel quantum phenomena arise due to interactions beyond the harmonic approximation, e.g. electron-phonon interactions (EPI)\cite{RevModPhys.89.015003} and phonon-phonon interactions (PPI)\cite{PhysRevB.84.094302}. These quantum phenomena is related to finite phonon lifetimes and phonon frequency shifts\cite{PhysRevB.104.224304}. Fundamental properties of crystals, such as electrical resistivity\cite{PhysRevB.54.16591} and superconductivity\cite{PhysRevB.12.905}, are determined by EPI. Phonon-phonon interactions in materials arise from the inherent anharmonicity in the crystal lattice. They give rise to a finite value of lattice thermal conductivity ($\kappa_{\rm{ph}}$). In addition to this, a variety of novel phenomena, such as phonon drag\cite{ashcroft} and hydrodynamic phonon transport\cite{science.aav3548}, are also influenced by PPI.

Body-centered cubic (BCC) transition metals have been a topic of great interest due to their interesting emergent and complex physical properties. In the field of lattice dynamics, high superconducting temperature ($T_c$) and electronic topological transitions (ETT)\cite{PhysRevB.92.134422} are the most well-known examples in V and Nb. The high $T_c$ of V (5.3 K\cite{PhysRev.85.85, PhysRevB.67.094509}) and Nb (9.25 K\cite{PhysRevLett.79.4262}) is directly related to EPI. These class of material are called BCS superconductors. Also, for the given class of materials $T_c$ depends on strength of EPI. ETT are directly associated with the variation in Fermi surface nesting and these transitions cause the anomalies in phase stability and elasticity of the material\cite{PhysRevB.75.064102}. Specifically, these novel phenomena occur at high pressure. The Kohn anomaly is one of the important anomalies in the lattice dynamics properties of BCC transition metals\cite{PhysRevLett.103.235501, computation6020029}. It is characterised by the strength of EPI. This anomaly affects the superconducting properties and lattice stability of the superconducting metals.

The lattice vibration in metals has been extensively studied for almost a half-century. Over the past two decades, density-functional theory (DFT) has become a very popular method for accurately calculating the thermophysical properties of a wide range of materials. Using the first principle approach, Luo \textit{et al.}\cite{pnas.0707377104} have studied the lattice dynamics of V under high pressure. They confirmed the complex phase transitions [BCC $\rightarrow$ RH$_1$($110.5^\circ$) $\rightarrow$ distorted-RH$_2$($108.2^\circ$) $\rightarrow$ BCC] with increasing pressure up to 400 GPa. Also, Bosak \textit{et al.}\cite{Bosak} measured phonon dispersions of V using inelastic x-ray scattering (IXS). They found several phonon-dispersion anomalies. Moreover, Antonangeli \textit{et al.}\cite{Daniele} conducted experimental research to measure the phonon dispersion of V at high pressure using IXS. Their results clearly showed the anomalous high-pressure behaviour of the transverse acoustic mode along the (100) direction. Furthermore, Nakagawa et al.\cite{PhysRevLett.11.271} measured the phonon dispersion of Nb by means of IXS. They also found abnormal features in the phonon dispersion of Nb. In addition to this, at present, there are several theoretical studies of the lattice dynamics of V and Nb\cite{pnas.0707377104, PhysRevB.54.16487, PhysRevB.77.024110, PhysRevB.101.115119}. In their work, they have also compared their theoretical results with the experimentally measured phonon dispersions. All previous theoretical studies on the phonon dispersion of V and Nb mainly addressed either the temperature effect or the pressure effect. But the effect of variation in the supercell size on the phonon dispersion is ignored in V and Nb, which, however, might be important. For the experimentally unexplored new material, how can one determine the phononic properties of a solid just by seeing the mechanical stability of the compound? Also, what is the origin of the dip-like feature in phonon dispersion along $\Gamma$-H high-symmetry directions, and how does this \enquote{dip}-like behaviour change with increasing the supercell size in the same high-symmetry path? Therefore, it becomes necessary to reinvestigate the lattice dynamic properties of V and Nb with different supercell sizes. Along with lattice dynamic properties, the carrier lifetime is an essential parameter that governs the temperature-dependent transport behaviour of materials. The carrier lifetime in a material is influenced by three main factors: 1) sources of scattering or interaction, 2) the strength of these interactions, and 3) the type of interactions experienced by the charge carriers. Electron-electron interaction, EPI, and PPI are intrinsic sources of scattering mechanisms that can strongly impact their macroscopic properties. The quantification of these scattering mechanisms through lifetime calculation is a valuable tool to understand charge carrier transport. Thus, using the DFT method, phonon lifetime due to EPI and PPI calculations become inevitable.  

To address these questions, we explored the extent to which phonon dispersion changes with the variation in the supercell size. The variations are summarised and compared with the available experimental data. In addition to this, the phonon density of states and the phononic part of thermodynamical properties (specific heat at constant volume, Helmholtz free energy and entropy) of V and Nb are computed. Further, the temperature-dependent variation of phonon lifetimes due to EPI and PPI are calculated. The effect of temperature on phonon lifetime is also explored across different branches. Along with this, $\kappa_{\rm{ph}}$ is also computed. The phonon dispersion and thermophysical properties obtained using the first-principles approach give a good explanation of the available experimental data.

%***************************************** Computational  **************************************
\section{Computational details}
The first-principles calculations based on density functional theory (DFT) are carried out using WIEN2k package\cite{blaha2020wien2k}. The phonon dispersions and phonon density of states of BCC Vanadium (V) and Niobium (Nb) are calculated using augmented plane wave plus local orbitals (APW+lo) method with the WIEN2k package and the PHONOPY package\cite{togo} with the supercell size 2$\times$2$\times$2, 3$\times$3$\times$3 and 4$\times$4$\times$4. Local density approximation (LDA)\cite{Perdew} is used as exchange-correlation (XC) functional in these calculations. The space group and optimized lattice constant for V (Nb) are taken as Im$\bar{3}$m and 2.9296 (3.2578) \AA, respectively. The DFT based force calculations are performed over a \textit{k}-mesh size of 16$\times$16$\times$16. To ensure the accuracy of the interatomic forces, the force convergence limit is set to $10^{-5}$ Ry/Bohr.

The calculations of phonon lifetime due to electron-phonon interactions are carried out using Abinit code\cite{gonze2016recent}. These computations are performed using density functional perturbation theory method\cite{PhysRevB.43.7231}. The projector augmented wave (PAW)\cite{Bloachi} pseudopotential is adopted. These calculations are performed over a \textit{k}-mesh size of 36$\times$36$\times$36 and \textit{q}-mesh size of 6$\times$6$\times$6. The energy cutoff parameter is set to 30 Hartree. To calculate the lifetime of phonons and the lattice thermal conductivity due to phonon-phonon interactions, the PHONO3PY package\cite{PhysRevB.91.094306} is used. Initially, the forces on each atom of $4\times 4\times 4$ supercell are calculated using the PAW method of DFT as implemented in the Abinit code\cite{gonze2016recent}. These calculations are performed over a $k$-mesh size of $4\times 4\times 4$ with the convergence limit set to $5\times 10^{-8}$ Ha/Bohr for the self-consistent calculations. The anharmonic force constants are obtained from Abinit calculations under the supercell method with a finite displacement of 0.08 bohr in the PHONO3PY package\cite{PhysRevB.91.094306}. This anharmonic force constants is used to further compute the imaginary part of phonon self-energy (Im$\Sigma_{\rm{ph}}$). The phonon lifetime and lattice thermal conductivity are obtained using a $q$-mesh size of $32\times 32\times 32$. A real-space cutoff distance of 7.9 (8.8) Bohr is used to reduce the number of supercell calculations effectively in V (Nb).

\begin{figure}\label{Fig.super_compare}
\includegraphics[width=1.00\linewidth, height=8.0cm]{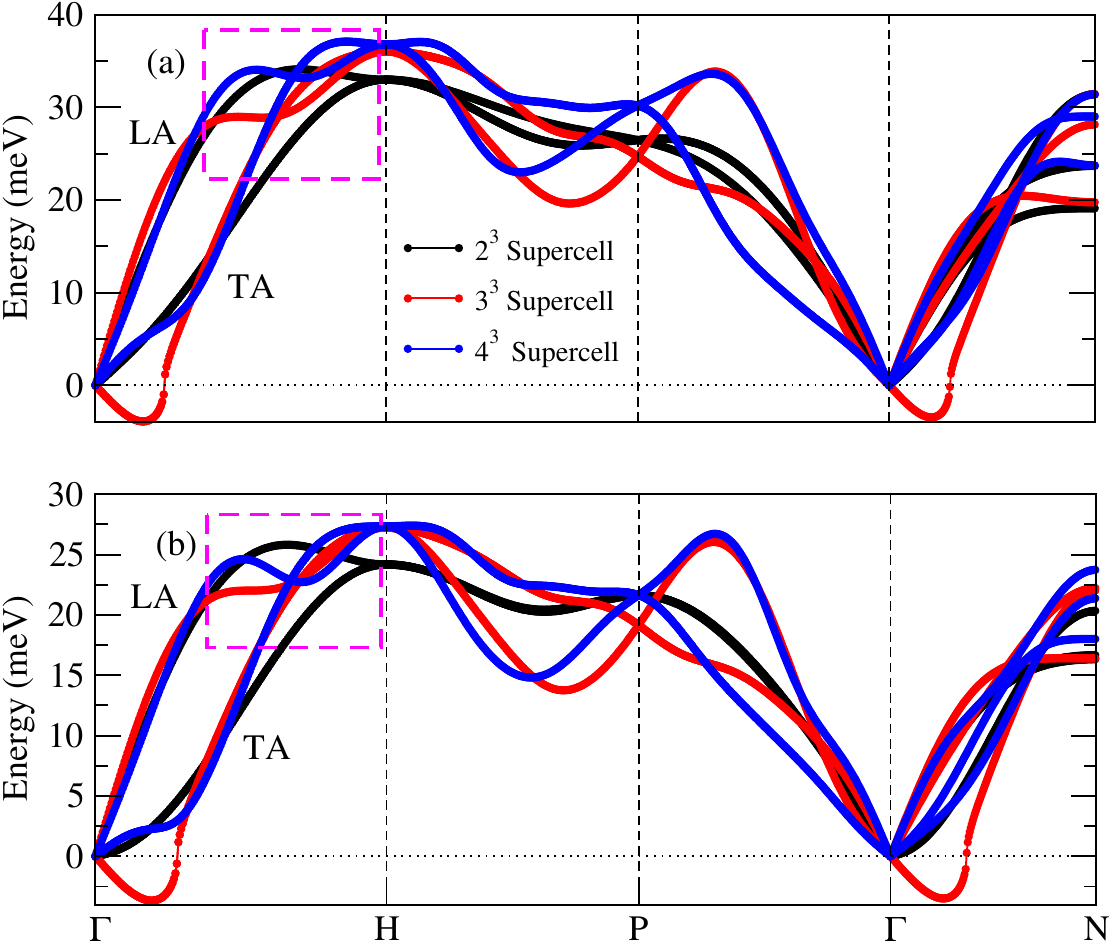} 
\caption{\label{Fig.super_compare}\small{Calculated phonon dispersion curve of the (a) V (b) Nb, using $2\times 2\times 2$, $3\times 3\times 3$ and $4\times 4\times 4$ supercell. }}
\end{figure}
 
\begin{figure}\label{Fig.exp_compare}
\includegraphics[width=1.00\linewidth, height=8.0cm]{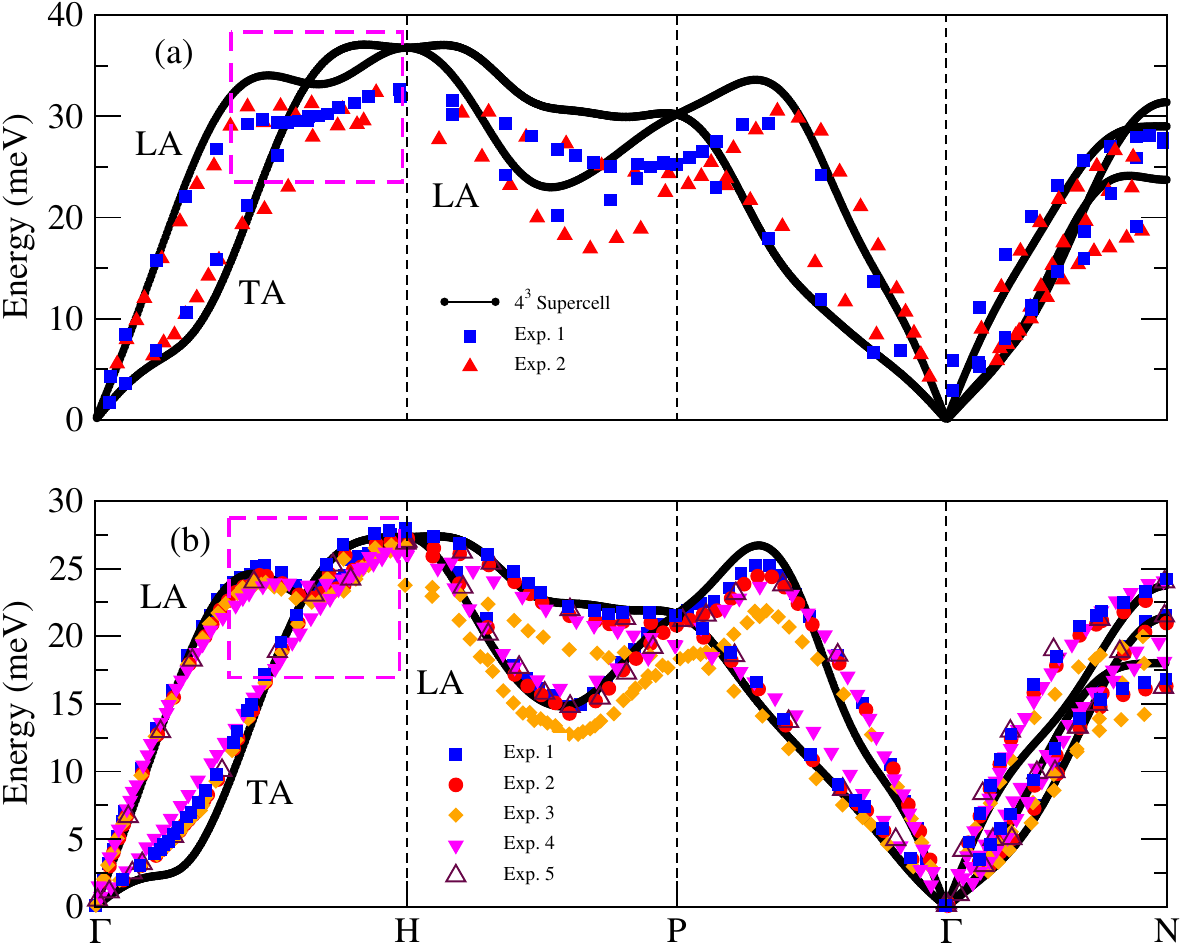}
\caption{\label{Fig.exp_compare}\small{Calculated phonon dispersion curve of the (a) V (b) Nb, using $4\times 4\times 4$ supercell. The splitted data-points represents experimental results of V [Exp.$\rightarrow$ 1\cite{Bosak} \& 2\cite{PhysRevB.1.3913}] and Nb [Exp.$\rightarrow$ 1\cite{Bosak}, 2\cite{PhysRevLett.11.271}, 3\cite{PhysRev.171.727}, 4\cite{PhysRevB.66.064303} \& 5\cite{Guthoff_1994}] }}
\end{figure}

%***************************************** Result and discussion **************************************  
\section{Results and Discussion}
\subsection{\label{sec:level2}Phonon dispersion and phonon density of states }
In order to test the sensitivity of the phonon dispersion with respect to $k$-mesh, the calculations of phonon at 2$\times$2$\times$2, 3$\times$3$\times$3 and 4$\times$4$\times$4 supercell with different sizes of $k$-mesh are carried out. In these calculations, it is found that the phonon dispersion curves are highly sensitive to the size of $k$-mesh used. Keeping 6$\times$6$\times$6 $k$-mesh in the phonon calculation of V for 2$\times$2$\times$2 supercell, imaginary frequencies are found along $\Gamma$-N high-symmetry direction. However, when the $k$-mesh is increased to 8$\times$8$\times$8, the observed imaginary frequencies are completely disapperaed. On further increasing the $k$-mesh to 10$\times$10$\times$10, no significant improvement can be seen. However, increasing the supercell size to 3$\times$3$\times$3, imaginary frequencies are obtained along the high-symmetry directions $\Gamma$-H-P-$\Gamma$-N corresponding to 6$\times$6$\times$6 $k$-mesh. In an attempt to remedy the unphysical mechanical instability obtained for phonons, $k$-mesh size in the calculations is increased upto 20$\times$20$\times$20. But the imaginary frequencies are not completely disappeared, even by taking such a large $k$-mesh size. At this $k$-mesh size the imaginary frequencies are observed along the small region of $\Gamma$-H and $\Gamma$-N high-symmetry directions (solid red curve of Fig. \ref{Fig.super_compare} (a)). Moving futher, for 4$\times$4$\times$4 supercell calculations, a starting $k$-mesh size of 4$\times$4$\times$4 is taken. At this $k$-mesh size, the imaginary frequencies are found along $\Gamma$-H-P-$\Gamma$-N high-symmetry directions. Increasing the $k$-mesh size further to 12$\times$12$\times$12 corresponding to supercell 4$\times$4$\times$4, small imaginary frequencies are observed in the vicinity of the $\Gamma$ point. To get the complete disappearance of the imaginary frequencies, a further increase in $k$-mesh size to 14$\times$14$\times$14 is used. At this $k$-mesh size, instabilities are completely removed. To check whether this is accidental, the $k$-mesh size was increased to 16$\times$16$\times$16, and it was found that there is no significant change in the phonon dispersions. Thus, for studying the phononic properties of V, optimized size of $k$-mesh is chosen as 16$\times$16$\times$16. A similar $k$-mesh convergence study is done on Nb for supercell sizes of 2$\times$2$\times$2, 3$\times$3$\times$3 and 4$\times$4$\times$4. It is found to have almost same convergence result as V except for the supercell size of 2$\times$2$\times$2. In 2$\times$2$\times$2 supercell size of Nb, it is found that the imaginary frequencies completely disappear at 10$\times$10$\times$10 $k$-mesh size. The above discussion reveals that before going to further calculations of the phononic properties of materials, $k$-mesh sensitivity should be checked. Along with $k$-mesh convergence, the qualitative behaviour of the phonon dispersion curves of V and Nb due to the variation in the supercell size is also studied.

The phonon dispersions obtained for the longitudinal acoustic (LA) and transverse acoustic (TA) modes using DFT based calculations are shown in Fig. \ref{Fig.super_compare}, along the high-symmetry directions $\Gamma$-H-P-$\Gamma$-N in the first Brillouin zone (BZ). The variations in the phonon dispersion of V and Nb due to 2$\times$2$\times$2, 3$\times$3$\times$3 and 4$\times$4$\times$4 supercells are shown in figure. From the figure, it is seen that both elements are mechanically stable at 0 K for supercell size 2$\times$2$\times$2. However, with increasing the supercell size to 3$\times$3$\times$3, imaginary frequencies are found in the small region of $\Gamma$-H and $\Gamma$-N high symmetric directions. On further increase in supercell size to 4$\times$4$\times$4, imaginary frequencies completely disappear in both elements. In addition to this, the maximum phonon energy of the TA phonon mode in V (Nb) is found to be 32.91 (24.26), 36.12 (27.34) and 36.45 (27.34) meV corresponding to supercell size of 2$\times$2$\times$2, 3$\times$3$\times$3 and 4$\times$4$\times$4, respectively. Furthermore, in a 4$\times$4$\times$4 (3$\times$3$\times$3) supercell of V a small kink (very small kink) is observed in the vicinity of the $\Gamma$-point along $\Gamma$-H high symmetric directions (TA mode of solid blue and black curve in Fig. \ref{Fig.super_compare} (a)). However, in the case of Nb, the size of kink is bigger than the V corresponding to 4$\times$4$\times$4 supercell. Also, it is bigger in comparison to the 3$\times$3$\times$3 supercell, which is shown in Fig. \ref{Fig.super_compare} (b). This kink-like feature, which coincides with the Kohn anomaly\cite{PhysRevLett.2.393}, was also observed in the phonon dispersions of V calculated using the stochastically initialized temperature-dependent effective potential method by Yang \textit{et al.}\cite{PhysRevB.101.094305}. A kink-like feature in the phonon dispersion typically refers to abrupt changes in the slope of the dispersion. This can occur due to various physical phenomena, such as lattice anomalies\cite{PhysRevLett.108.045502, Daniele}, phase transitions\cite{PhysRevB.83.054101}, or EPI\cite{PhysRevB.37.10674, PhysRevB.82.195121}. This feature, associated with strong electron-phonon coupling, can be of significance in the context of superconductivity. Moreover, it plays a crucial role in the formation of Cooper pairs\cite{Ishida2016}. It is important to mention here that the phonon calculation is performed at 0 K using the DFT method. However, the available experimental data is at room temperature. Hence, it is obvious that the experimental result will always differ from the theoretical result at two different temperatures. Moreover, the volume of the primitive unit cell of V and Nb will change with temperature, resulting in a change in the lattice parameter. Therefore, phonon dispersion is expected to change.

A noteworthy \enquote{dip}-like feature in the LA phonon mode along the $\Gamma$-H high symmetric path is observed for V and Nb in the case of supercell size 4$\times$4$\times$4. This feature is indicated by the dotted square box in both figures (Fig. \ref{Fig.super_compare} (a) \& (b)). Also, in supercell size 3$\times$3$\times$3, a small dip is also seen in the LA phonon mode. But a small instability occurs in this supercell size along $\Gamma$-H and $\Gamma$-N high symmetric directions, which is spurious. However, in the supercell size 2$\times$2$\times$2, \enquote{dip}-like feature is absent in both elements. The origin of the \enquote{dip}-like feature in the different sizes of supercells is due to long-range interactions, which play a crucial role in the phonon dispersion curve\cite{ashcroft}. This \enquote{dip}-like feature in the phonon dispersion is termed as soft mode. From the obtained results, it is found that the soft mode mainly depends on the size of the supercell. We therefore conclude that the supercell size 4$\times$4$\times$4 is more accurate for the calculation of phononic properties of V and Nb. Along with the change in dispersion with the variation in the supercell size, the phonon dispersion curve at supercell size 4$\times$4$\times$4 is also compared with available experimental data\cite{PhysRevB.1.3913, Bosak, PhysRevLett.11.271, PhysRev.171.727, PhysRevB.66.064303, Guthoff_1994}.

\begin{figure}\label{Fig.DOS}
\includegraphics[width=1.00\linewidth, height=8.0cm]{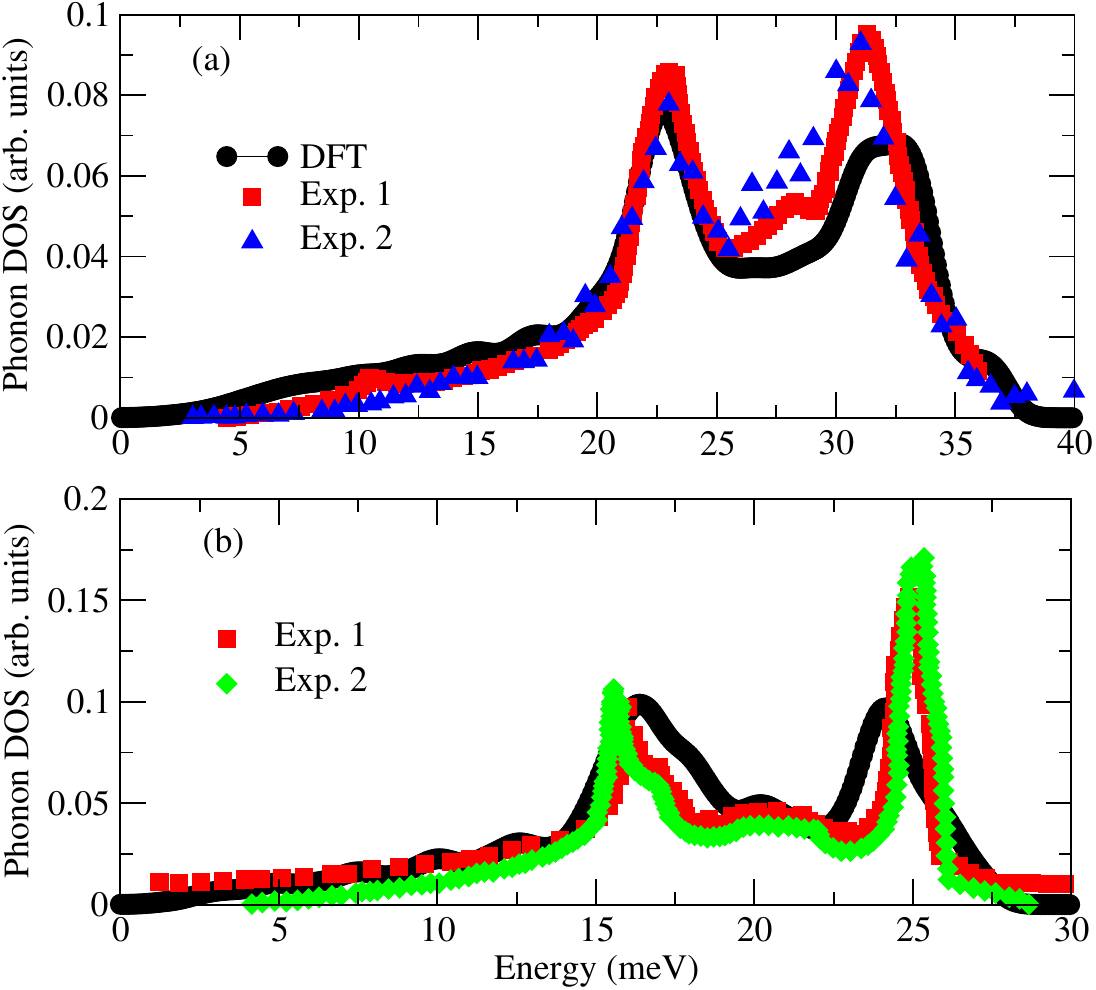}
\caption{\label{Fig.DOS}\small{Calculated phonon density of states (DOS) of the (a) V (b) Nb, using $4\times 4\times 4$ supercell. The splitted data-points represents experimental results of V [Exp.$\rightarrow$ 1\cite{Sears} \& 2\cite{PhysRevB.65.014303}] and Nb [Exp.$\rightarrow$ 1\cite{Sears} \& 2\cite{Guthoff_1994}] }}
\end{figure}

\begin{figure}\label{Fig.thermal}
\includegraphics[width=1.00\linewidth, height=8.0cm]{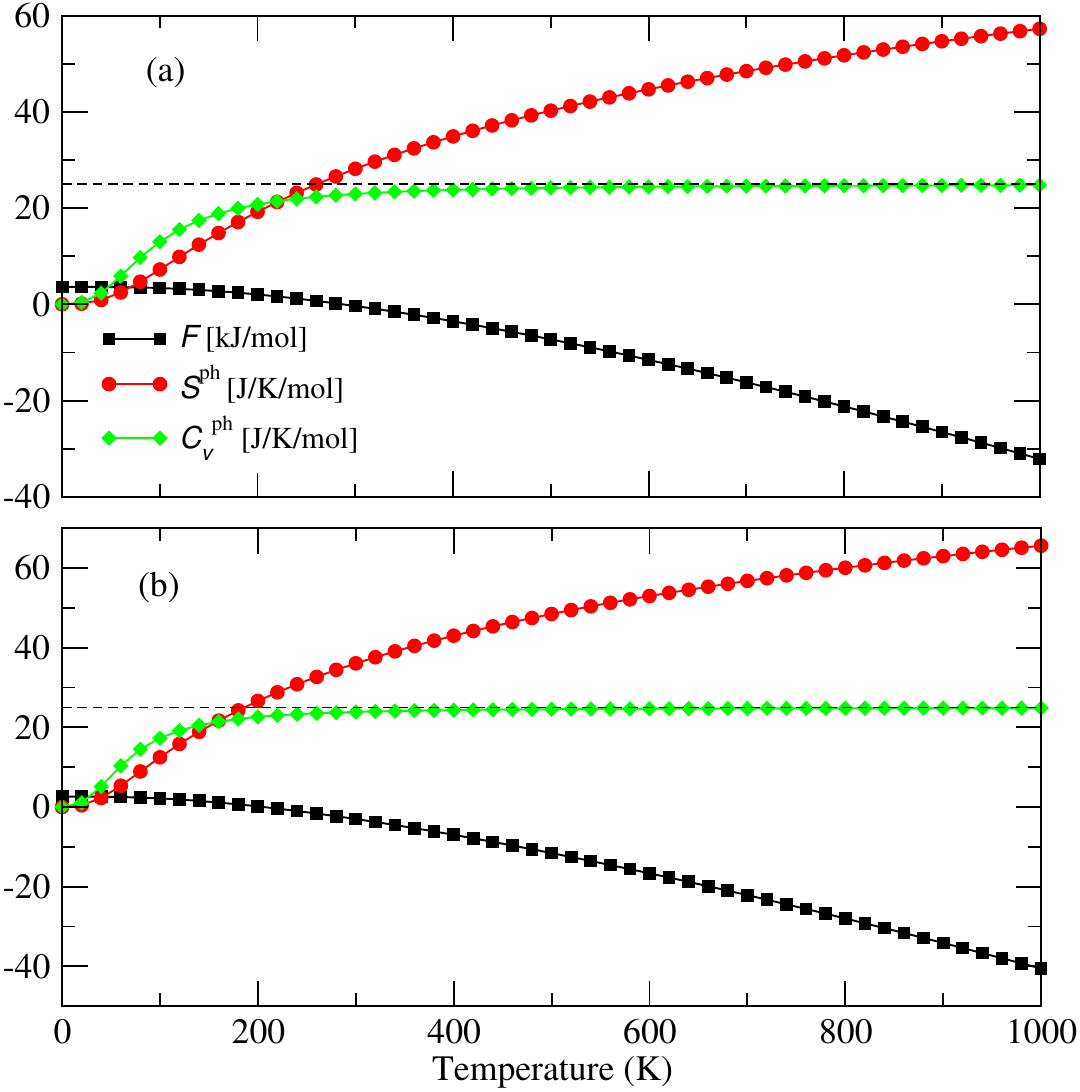}
\caption{\label{Fig.thermal}\small{Phononic part of Helmholtz free energy ($F$), entropy ($S^{\rm{ph}}$) and specific heat at constant volume ($C_v^{\rm{ph}}$) (a) V (b) Nb.}}
\end{figure}

In order to have a better picture of the lattice dynamics of solids, phonon density of states (DOS) is crucial because it determines several material properties such as thermal and phononic transport, thermodynamics, and structural stability. Fig. \ref{Fig.DOS} (a) shows the phonon DOS of V along with the available experimental data\cite{Sears, PhysRevB.65.014303}. In the figure, a small peak around 10 meV is seen in the result of Sears \textit{et al.}, which is not found in our data. However, no such small feature has been reported in the available experimental data\cite{PhysRevB.65.014303}, except for the results of Sears \textit{et al.}. This indicates that the small feature obtained at this energy is uncertain. A small hump around 29 meV is also observed in Sears \textit{et al.}'s data, but we do not find such a hump. In addition to this, it is also seen that the height of the peak around 33 meV is less than in both reported experimental data\cite{Sears, PhysRevB.65.014303}. This may be due to the fact that the present calculation is at 0 K while the available experimental data is at room temperature. The maximum frequency is found to be 36.7 meV. Also, between 20 and 37 meV, three distinct peaks in the DOS are observed at 23.05, 31, and 33.64 meV. It is important to mention here that in phonon dispersions, the highest energy band is at 36.77 meV, which is almost the same as the energy corresponding to DOS. Hence, the calculated phonon DOS of V is in overall good agreement with the available experimental data\cite{Sears, PhysRevB.65.014303}. Fig. \ref{Fig.DOS} (b) depicts the phonon DOS of Nb. The obtained phonon DOS are also compared with the available experimental data\cite{Sears, Guthoff_1994}. From the figure, it is seen that the obtained phonon DOS is qualitatively similar to the experimental results. The figure also shows that the first peak is present at 16.31 meV, which is almost the same energy as experimentally reported. Similarly, the second peak, which is present at 24.04 meV, is also well in agreement with the experimental peak. In addition to this, the maximum energy is found to be 27.5 meV, which is almost the same as the highest phonon branch energy (27.34 meV). It is also important to mention here that, in the obtained results, the height of the second peak is less than the experimentally reported height. This is possibly because present calculations are performed at 0 K. Hence, such deviations are expected at 0 K.
  
%%%%%%%%%%%%%%%%%%%%%%%%%%%%%%%% Cv table of V
\begin{table}\label{tabVCv}
\caption{\label{tabVCv}
\small{Calculated values of phonon heat capacity ($C_v^{\rm{ph}}$) of V at 300 K along with the extracted values of phonon heat capacity ($C^{\rm{ph}}_{{v}_{\rm{extracted}}}$) from the corresponding experiments ($C_{v}^{\rm{experiment}}$).}}

\begin{ruledtabular}
\begin{tabular}{ cccc } 
\textrm{Our Work}&
\textrm{{$C^{\rm{ph}}_{{v}_{\rm{extracted}}}$}}&
\textrm{{$C_{v}^{\rm{experiment}}$}}\\
\textrm{(J/K/mol)}&
\textrm{{(J/K/mol)}}&
\textrm{{(J/K/mol)}}\\
\colrule
\multirow{3}{5em}{22.94} & 22.10 & 24.89\cite{hultgren1973selected} \\ 
& 22.12 & 24.90\cite{chase1982janaf} \\ 
& 21.83 & 24.61\cite{Desai1986} \\ 
\end{tabular}
\end{ruledtabular}
\end{table}

%------------------------------------------Thermal properties------------------------- 

\subsection{\label{sec:level2}Thermodynamical properties} 
The calculations of temperature-dependent phononic part of thermodynamical properties such as Helmholtz free energy ($F$), entropy ($S^{\rm{ph}}$) and heat capacity at constant volume ($C_v^{\rm{ph}}$) are performed within the temperature range of $0-1000$ K using relations as follows\cite{dove1993introduction},

\begin{equation}\label{eq.F}
F=\frac{1}{2}\sum_{\textbf{q}j}\hbar\omega_{\textbf{q}j}+k_BT\sum_{\textbf{q}j}\ln[1-\exp(-\hbar\omega_{\textbf{q}j}/k_BT)]
\end{equation}

\begin{equation}\label{eq.C}
C_v^{\rm{ph}}=\sum_{\textbf{q}j}k_B\left(\frac{\hbar\omega_{\textbf{q}j}}{k_BT}\right)^2\frac{\exp(\hbar\omega_{\textbf{q}j}/k_BT)}{[\exp(\hbar\omega_{\textbf{q}j}/k_BT)-1]^2}
\end{equation}
and
\begin{equation}\label{eq.S}
S^{\rm{ph}}=\frac{1}{2T}\sum_{\textbf{q}j}\hbar\omega_{\textbf{q}j}\coth\left(\frac{\hbar\omega_{\textbf{q}j}}{2k_BT}\right)-k_B\sum_{\textbf{q}j}\ln\left[ \sinh\left(\frac{\hbar\omega_{\textbf{q}j}}{2k_BT}\right)\right] 
\end{equation}
where $k_B$, $\hbar$ and $T$ represent Boltzmann constant, the reduced Planck’s constant and temperature, respectively. In addition to this, $\omega_{\textbf{q},j}$ denotes the phonon frequency of mode $\left\lbrace\textbf{q}j\right\rbrace$, where $\textbf{q}$ is wave vector and $j$ is branch index. The calculated values $F$, $S^{\rm{ph}}$ and $C_v^{\rm{ph}}$ corresponding to the temperature range $0-1000$ K is shown in Fig. \ref{Fig.thermal}. From the figure, it is clear that value of $F$ is monotonically decreasing with the rise in $T$ in the given temperature range. In the case of V (Nb), it is found that the values of $F$ is positive within the temperature range $0-287$ (206) K while, for $T>287$ (206) K, the values of $F$ are negative. For V (Nb), the value of $F$ is found to be 0 at $\sim$ 288 (207) K, which gets further decreased to -32.13 (-40.55) kJ/mol at 1000 K. It is also found that at 0 K, the magnitude of $F$ is non zero due to quantum fluctuation. According to quantum mechanics, there is always a vibrational energy at 0 K. This energy is called zero point energy.

%%%%%%%%%%%%%%%%%%%%%%%%%%%%%%%% Cv table of Nb
\begin{table}\label{tabNbCv}
\caption{\label{tabNbCv}
\small{Calculated values of phonon heat capacity ($C_v^{\rm{ph}}$) of Nb at 300 K along with the extracted values of phonon heat capacity ($C^{\rm{ph}}_{{v}_{\rm{extracted}}}$) from the corresponding experiments ($C_{v}^{\rm{experiment}}$).}}

\begin{ruledtabular}
\begin{tabular}{ cccc } 
%\hline
\textrm{Our Work}&
\textrm{{$C^{\rm{ph}}_{{v}_{\rm{extracted}}}$}}&
\textrm{{$C_{v}^{\rm{experiment}}$}}\\
\textrm{(J/K/mol)}&
\textrm{{(J/K/mol)}}&
\textrm{{(J/K/mol)}}\\
\colrule
\multirow{3}{5em}{23.86} & 22.26 & 24.60\cite{hultgren1973selected} \\ 
& 22.35 & 24.69\cite{chase1982janaf} \\ 
& 22.25 & 24.59\cite{kirillin1965thermodynamic} \\ 
%\hline
\end{tabular}
\end{ruledtabular}
\end{table}

Let us now focus on the temperature-dependent phonon heat capacity at constant volume, $C_v^{\rm{ph}}$. Indeed, heat capacity is a thermodynamic property that is typically measured experimentally. The calculated phonon heat capacities are shown in Fig. \ref{Fig.thermal}. The dashed line in the figure represents the classical Dulong and Petit’s limit. From Fig. \ref{Fig.thermal} (a), it is seen that the values of $C_v^{\rm{ph}}$ increase with the rise in $T$. The value of $C_v^{\rm{ph}}$ of V at 300 K is found to be 22.94 J/K/mol, which gets nearly constant when $T>600$ K. It is known that the heat capacity of materials, $C_v$, is the sum of the electronic component ($C_v^{\rm{el}}$), phononic component ($C_v^{\rm{ph}}$), and magnonic component ($C_v^{\rm{magn}}$), i.e., $C_v=C_v^{\rm{el}}+C_v^{\rm{ph}}+C_v^{\rm{magn}}$\cite{grimvall1999thermophysical}. Since V and Nb are non-magnetic systems, the dominant heat capacity should come only from lattice vibrations and electronic excitations. In this case, the magnonic component would be negligible or effectively zero. For common metals, electronic heat capacity, $C_v^{\rm{el}}$ is defined as\cite{grimvall1999thermophysical},

\begin{equation}\label{eq.Cel}
C_v^{\rm{el}}=\gamma T
\end{equation} 
 
where $\gamma$ represents the electronic heat capacity coefficient. The reported value of $\gamma$ for V is 9.26 mJ/mol/K$^2$\cite{kittel2005introduction}. According to Eq. \ref{eq.Cel}, the electronic heat capacity $C_v^{\rm{el}}$ at 300 K should be about 2.78 J/K/mol. Moreover, the experimentally reported total heat capacities at room temperature are presented in Table \ref{tabVCv}. Therefore, the extracted phononic heat capacities ($C_v^{\rm{ph}}=C_v-C_v^{\rm{el}}$) from the corresponding experimental data at 300 K are 22.10, 22.12 and 21.83 J/K/mol, which are mentioned in table. The calculated value of $C_v^{\rm{ph}}$ gets nicely matched with that of the extracted values of phonon heat capacity. Like V, the temperature-dependent phonon heat capacity of Nb is also calculated. The results are depicted in Fig. \ref{Fig.thermal} (b). The value of $C_v^{\rm{ph}}$ is found to be 23.86 J/K/mol at 300 K, which gets nearly constant when $T>400$ K. Since the reported value of $\gamma$ for Nb is 7.79 mJ/mol/K$^2$\cite{kittel2005introduction}. Hence, the electronic heat capacity $C_v^{\rm{el}}$ at 300 K should be about 2.34 J/K/mol. Furthermore, as reported in experimental studies the total heat capacities at 300 K are mentioned in Table \ref{tabNbCv}. Therefore, the extracted phononic heat capacities from the corresponding experimental data\cite{hultgren1973selected, chase1982janaf, kirillin1965thermodynamic} at 300 K are 22.26, 22.35 and 22.25 J/K/mol. Also, the obtained results at 300 K are given in table, along with the extracted values of $C_v^{\rm{ph}}$ corresponding to experimental results\cite{hultgren1973selected, chase1982janaf, kirillin1965thermodynamic}. The calculated value of $C_v^{\rm{ph}}$ is in good agreement with that of the extracted values.

\begin{table}\label{tabVS}
\caption{\label{tabVS}%
\small{Calculated values of phonon entropy ($S^{\rm{ph}}$) of V at 300 K along with the extracted values of phonon entropy ($S^{\rm{ph}}_{\rm{extracted}}$) from the corresponding experiments ($S_{\rm{experiment}}$).}}

\begin{ruledtabular}
\begin{tabular}{ cccc } 
\textrm{Our Work}&
\textrm{{$S^{\rm{ph}}_{\rm{extracted}}$}}&
\textrm{{$S_{\rm{experiment}}$}}\\
\textrm{(J/K/mol)}&
\textrm{{(J/K/mol)}}&
\textrm{{(J/K/mol)}}\\
\colrule
\multirow{3}{5em}{28.16} & 28.11 & 30.89\cite{smith1981v} \\ 
& 27.01 & 29.79\cite{10.1007/s11669-016-0514-7} \\ 
& 26.93 & 29.71\cite{Desai1986} \\ 
\end{tabular}
\end{ruledtabular}
\end{table}

Apart from this, the calculation of temperature-dependent $S^{\rm{ph}}$ is also carried out using the DFT approach. Corresponding plot is shown in Fig. \ref{Fig.thermal}. From the Fig. \ref{Fig.thermal} (a), it is seen that the value of $S^{\rm{ph}}$ is increasing monotonically with the rise in $T$. For V, the magnitude of $S^{\rm{ph}}$ is found to be 28.16 J/K/mol at 300 K, which gets further increased to 57.26 J/K/mol at 1000 K. Similar to heat capacity, the entropy ($S$) of materials is influenced by the electronic ($S^{\rm{el}}$) and phononic ($S^{\rm{ph}}$) components\cite{grimvall1999thermophysical}. According to the general thermodynamic relation, the electronic entropy, $S^{\rm{el}}$ is defined as\cite{grimvall1999thermophysical}, 

\begin{equation}\label{eq.Sel}
S^{\rm{el}} (T)=\int_0^T \frac{C_v^{\rm{el}}(T^\prime)}{T^\prime} \,\mathrm {d} T^\prime
\end{equation} 

Thus, $S^{\rm{el}} (T)=C_v^{\rm{el}}(T)$. Since we have already calculated the electronic heat capacity $C_v^{\rm{el}}$ of V at 300 K, which is 2.78 J/K/mol. Moreover, the experimentally reported total entropies at room temperature are given in Table \ref{tabVS}. Therefore, the extracted phonon entropies ($S^{\rm{ph}}=S-S^{\rm{el}}$) from the corresponding experimental data at 300 K are 28.11, 27.01 and 26.93 J/K/mol, which are mentioned in table. The computed value of $S^{\rm{ph}}$ (28.16 J/K/mol) is good agreement with that of the extracted values of phonon entropy. We now focus on the temperature-dependent phonon entropy of Nb. The results are depicted in Fig. \ref{Fig.thermal} (b). The value of $S^{\rm{ph}}$ is found to be 36.06 J/K/mol at 300 K, which gets further increased to 65.59 J/K/mol at 1000 K. At 300 K, the electronic part of entropy is calculated as 2.34 J/K/mol. In addition to this, the experimentally reported total entropies at room temperature are presented in table \ref{tabNbS}. Thus, the extracted phonon entropies from the corresponding experiments at 300 K are 34.12, 34.12 and 34.06 J/K/mol. The extracted results are given in table. It is seen from the table that the extracted values are close to our calculated value of $S^{\rm{ph}}$ (36.06 J/K/mol).

Finally, we note that our phonon heat capacity and phonon entropy results for V and Nb at 300 K compare well with the extracted phonon entropies from different reported experimental data.

%------------------------------------------Phonon Lifetime-------------------------
\begin{figure}\label{Fig.V_tau}
\includegraphics[width=1.00\linewidth, height=9.0cm]{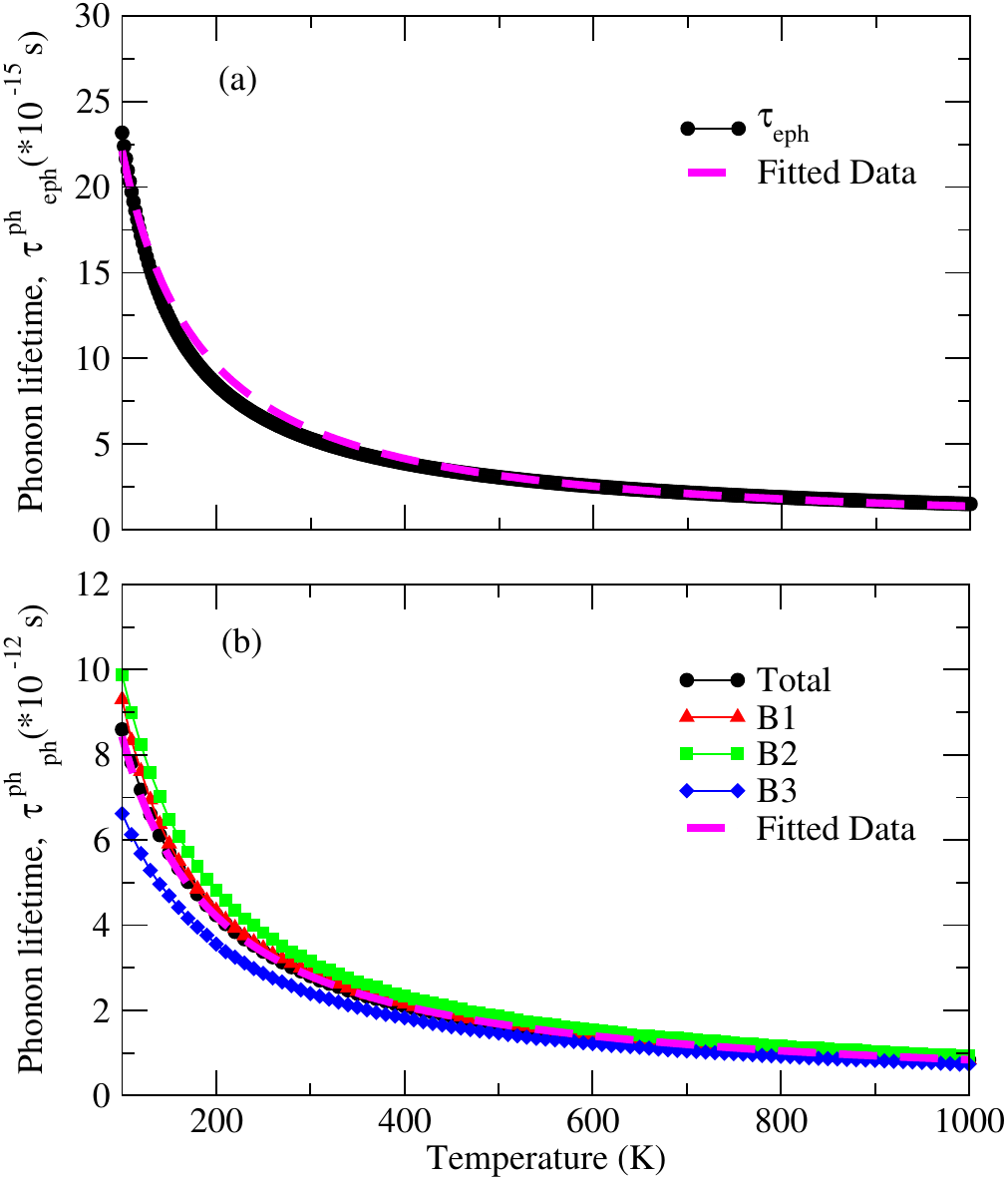}
\caption{\label{Fig.V_tau}\small{Phonon lifetime of V as a function of temperature (a) The phonon lifetime due to electron-phonon interaction ($\tau_{\rm{eph}}^{\rm{ph}}$). (b) The phonon lifetime due to phonon-phonon interaction ($\tau_{\rm{ph}}^{\rm{ph}}$) for total and three phonon branches. The magenta line represents the fitted value of phonon lifetime.}}
\end{figure}

\begin{figure}\label{Fig.Nb_tau}
\includegraphics[width=1.00\linewidth, height=9.0cm]{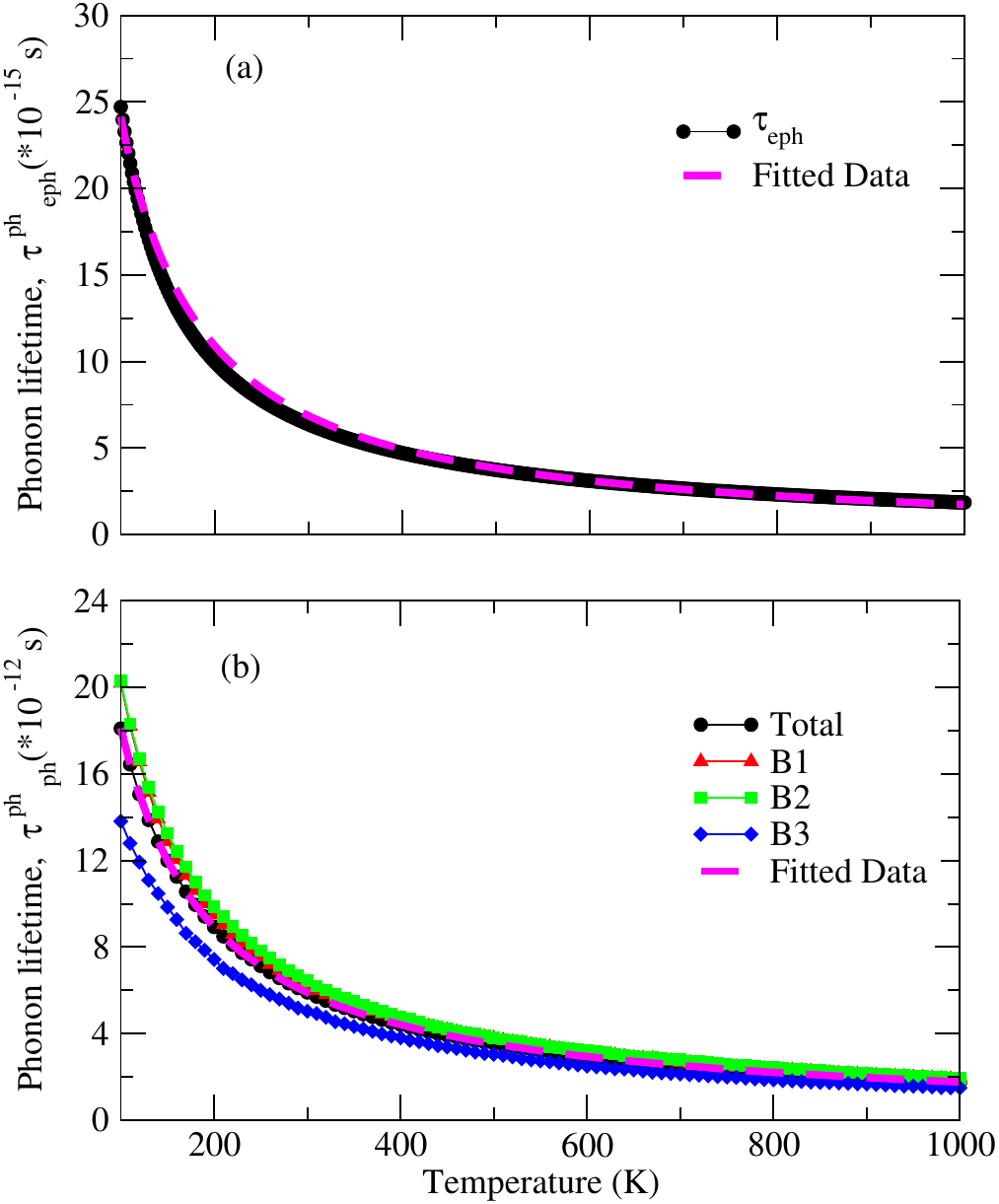}
\caption{\label{Fig.Nb_tau}\small{Phonon lifetime of Nb as a function of temperature (a) The phonon lifetime due to electron-phonon interaction ($\tau_{\rm{eph}}^{\rm{ph}}$). (b) The phonon lifetime due to phonon-phonon interaction ($\tau_{\rm{ph}}^{\rm{ph}}$) for total and three phonon branches. The magenta line represents the fitted value of phonon lifetime.}}
\end{figure}

\subsection{\label{sec:level2}Phonon lifetime}

For understanding the phononic transport properties such as thermal conductivity require knowledge of anharmonic vibrational properties of the material, the phonon lifetime is an important parameter. The major scattering mechanisms for the phonons are phonon-phonon interaction (PPI), electron-phonon interaction (EPI), and phonon-defect interaction, which determine the phonon lifetime of a material. In this section, the temperature-dependent phonon lifetime due to the EPI and PPI of V and Nb is calculated within temperature range 100-1000 K. In Fig. \ref{Fig.V_tau}, the calculated phonon lifetimes due to EPI ($\tau_{\rm{eph}}^{\rm{ph}}$) and PPI ($\tau_{\rm{ph}}^{\rm{ph}}$) of V are shown against temperature. From the Fig. \ref{Fig.V_tau} (a), it is  found that the values of $\tau_{\rm{eph}}^{\rm{ph}}$ decreases with the rise in $T$. In order to understand the scattering rate ($\tau^{-1}$) with temperature, a curve is fitted over the theoretical data. The form of the curve is as follows:
\begin{equation}\label{eq.Fit}
\tau(T)=\frac{a}{T^b}
\end{equation}

\begin{table}\label{tabNbS}
\caption{\label{tabNbS}%
\small{Calculated values of phonon entropy ($S^{\rm{ph}}$) of Nb at 300 K along with the extracted values of phonon entropy ($S^{\rm{ph}}_{\rm{extracted}}$) from the corresponding experiments ($S_{\rm{experiment}}$).}}

\begin{ruledtabular}
\begin{tabular}{ cccc } 
\textrm{Our Work}&
\textrm{{$S^{\rm{ph}}_{\rm{extracted}}$}}&
\textrm{{$S_{\rm{experiment}}$}}\\
\textrm{(J/K/mol)}&
\textrm{{(J/K/mol)}}&
\textrm{{(J/K/mol)}}\\
\colrule
\multirow{3}{5em}{36.06} & 34.12 & 36.46\cite{kirillin1965thermodynamic} \\ 
& 34.12 & 36.46\cite{chase1982janaf} \\ 
& 34.06 & 36.40\cite{hultgren1973selected} \\ 
\end{tabular}
\end{ruledtabular}
\end{table}

\begin{figure}\label{Fig.kappa_V}
\includegraphics[width=0.90\linewidth, height=5.0cm]{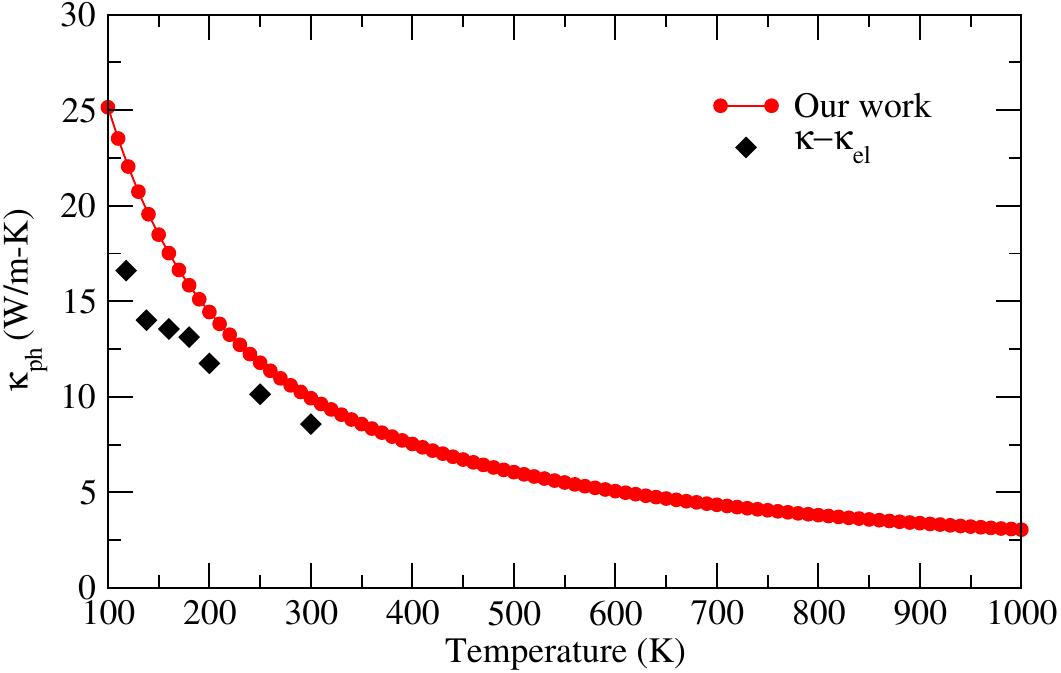}
\caption{\label{Fig.kappa_V}\small{Lattice thermal conductivity ($\kappa_{\rm{ph}}$) of V as a function of temperature. The splitted data points represent the extracted $\kappa_{\rm{ph}}$, which is obtained by subtracting the electronic contribution ($\kappa_{\rm{el}}$) from experimental thermal conductivity ($\kappa$) data.}}
\end{figure}

Here, for the EPI, the values of $a$ and $b$ are found to be $5664.9$ unit and $1.20$, respectively. From the fitted data, it is found that $\tau$ is proportional to $1/T^{1.2}$ within the temperature range of 100-1000 K, which is in a good match with the calculated values of $\tau$. The magnitude of $\tau_{\rm{eph}}^{\rm{ph}}$ is found to be $23.16\times 10^{-15}$ seconds (s) at $100$ K, which gets decreased to $1.51\times 10^{-15}$ s at 1000 K. The value of $\tau_{\rm{eph}}^{\rm{ph}}$ is reducing as we go from 100 K to 1000 K.  Along with this, $\tau_{\rm{ph}}^{\rm{ph}}$ is also calculated. Corresponding plot is shown in Fig. \ref{Fig.V_tau} (b). The phonon lifetime is calculated by taking the weight average over $q$-point within the same temperature range. A higher lifetime branch will make a lesser contribution to the phonon scattering mechanisms. From the figure, it is seen that the acoustic mode B2 (B3) has the highest (lowest) lifetime, indicating the lowest (highest) scattering. The lifetime of all the acoustic modes decreases with the rise in temperature. Further, the overall phonon lifetime is also calculated by averaging the lifetime of each acoustic mode at each temperature, as shown in Fig. \ref{Fig.V_tau} (b). The acoustic mode B2 show a higher lifetime as compared to the B1 and B2 mode. But the lifetimes of all acoustic modes come closer with the rise in temperature. The phonon lifetimes of B1, B2 and B3 modes are calculated as $9.29\times 10^{-12}$ ($0.84\times 10^{-12}$), $ 9.87\times 10^{-12}$ ($ 0.93\times 10^{-12}$) and $6.62\times 10^{-12}$ ($0.73\times 10^{-12}$) s at 100 (1000) K, respectively. Also, the total phonon lifetimes are found to be $8.59\times 10^{-12}$ and $0.83\times 10^{-12}$ s at 100 and 1000 K, respectively. To understand the phonon lifetime behaviour against temperature, Eq. \ref{eq.Fit} is also fitted over theoretical data. For this case, the values of $a$ and $b$ are found to be $841.48$ unit and $\sim 1$, respectively. From the fitted data, it is found that $\tau$ is proportional to $1/T$ within the temperature range of 100-1000 K. A similar phonon lifetime study is done on Nb. It is found to have almost the same qualitative features in its lifetime as V. From the Fig. \ref{Fig.Nb_tau} (a), it is also found that the values of $\tau_{\rm{eph}}^{\rm{ph}}$ decreases as the temperature increases. The magnitude of $\tau_{\rm{eph}}^{\rm{ph}}$ is found to be $24.70\times 10^{-15}$ seconds (s) at $100$ K, which gets decreased to $1.85\times 10^{-15}$ s at 1000 K. In addition this, $\tau_{\rm{ph}}^{\rm{ph}}$ is also discussed. The phonon lifetimes of B1, B2 and B3 modes are calculated as $20.19\times 10^{-12}$ ($1.86\times 10^{-12}$), $20.28\times 10^{-12}$ ($1.93\times 10^{-12}$) and $13.81\times 10^{-12}$ ($1.491\times 10^{-12}$) s at 100 (1000) K, respectively. Also, the total phonon lifetimes are found to be $18.09\times 10^{-12}$ and $1.76\times 10^{-12}$ s at 100 and 1000 K, respectively. From the fitted data, it is found that $\tau_{\rm{eph}}^{\rm{ph}}$ ($\tau_{\rm{ph}}^{\rm{ph}}$) follows a $1/T^{1.14}$ ($1/T$) dependence, which is very close to our calculated results. The values of $a$ and $b$ is found to be $4645.7$ ($1768.7$) unit and $1.14$ ($\sim 1$) for $\tau_{\rm{eph}}^{\rm{ph}}$ ($\tau_{\rm{ph}}^{\rm{ph}}$), respectively. The above discussion reveals that the EPI is a major source of scattering in comparison to that of PPI in deciding phonon transport properties. Hence EPI cannot be ignored in determining transport properties for such compound. Along with phonon lifetime, the lattice thermal conductivity of V and Nb is also studied.

\subsection{\label{sec:level2}Lattice thermal conductivity}

Under the single-mode relaxation time (SMRT) approximation of linearized phonon Boltzmann equation, lattice thermal conductivity ($\kappa_{\rm{ph}}$) of V and Nb is calculated. It is important to mention here that in present calculations only PPI is considered. The phonon lifetime due to PPI of mode $\lambda$ is obtained from Im$\Sigma_{\rm{ph}}$. The imaginary part of phonon self-energy, $\Gamma_\lambda$ is calculated using the third order force constants by\cite{PhysRevB.97.224306}

\begin{multline}\label{eq.Img}
\Gamma_\lambda(\omega)=\frac{18\pi}{\hbar^2}\sum_{\lambda^\prime\lambda^{\prime\prime}}|\Phi_{\lambda\lambda^\prime\lambda^{\prime\prime}}|^2 \{(n_{\lambda^\prime}+ n_{\lambda^{\prime\prime}}+ 1)\\\times \delta(\omega -\omega_{\lambda^\prime} -\omega_{\lambda^{\prime\prime}}) + (n_{\lambda^\prime}-n_{\lambda^{\prime\prime}})\\ \times [\delta(\omega +\omega_{\lambda^\prime} -\omega_{\lambda^{\prime\prime}}) -\delta(\omega -\omega_{\lambda^\prime}+\omega_{\lambda^{\prime\prime}})]\}
\end{multline}

Here, $\Phi_{\lambda\lambda^\prime\lambda^{\prime\prime}}$ denotes strength of interaction between three phonon $\lambda$, $\lambda^\prime$ and $\lambda^{\prime\prime}$ in the scattering. $n_\lambda$ reperesents the Bose–Einstein distribution function at equilibrium and $\omega_{\lambda}$ is frequency of phonon mode. The phonon lifetime of mode $\lambda$ is obtained as $\tau_\lambda = 1/2\Gamma_\lambda(\omega)$\cite{PhysRevB.97.224306}. Here, $2\Gamma_\lambda(\omega)$ is the phonon linewidth of mode $\lambda$. Under the SMRT approximation, the $\kappa_{\rm{ph}}$ is obtained using linearized phonon Boltzmann equation as\cite{PhysRevB.97.224306, PhysRev.128.2589}:

\begin{equation}\label{eq.Kappa}
\kappa_{\rm{ph}}=\frac{1}{NV_0}\sum_{\lambda} C_\lambda \textbf{v}_\lambda\otimes\textbf{v}_\lambda\tau_\lambda^{\rm {SMRT}}
\end{equation}

where, $N$, $V_0$ and $\textbf{v}_\lambda$ denote the number of unit cells, volume of a unit cell and group velocity, respectively. $C_\lambda$ and $\tau_\lambda^{\rm {SMRT}}$ represent heat capacity and single-mode relaxation time of phonon mode $\lambda$. 

The obtained lattice thermal conductivity of V within the temperature range 100-1000 K is shown in Fig. \ref{Fig.kappa_V} along with the extracted $\kappa_{\rm{ph}}$. As shown in figure, $\kappa_{\rm{ph}}$ decreases with increasing temperature. At 300 K, the our calculated value of $\kappa_{\rm{ph}}$ is 9.93 W/m-K, while it reaches 3.05 W/m-K at 1000 K. However, there is no direct experimental result for comparison with our obtained results. As it is known, the thermal conductivity of materials, $\kappa$, is the sum of the electronic component ($\kappa_{\rm{el}}$) and phononic component ($\kappa_{\rm{ph}}$), i.e., $\kappa=\kappa_{\rm{el}}+\kappa_{\rm{ph}}$. According to the Wiedemann-Franz law, $\kappa_{\rm{el}}$ of a material is directly related to electrical conductivity\cite{ashcroft}. In case of common metals, the Wiedemann-Franz law is given by\cite{ashcroft},
\begin{equation}\label{eq.WFL}
\kappa_{\rm{el}}=\sigma T L
\end{equation}

where $\sigma$, $T$ and $L$ represent the electrical conductivity, temperature and Lorenz number ($2.44\times 10^{-8}$ W$\Omega$K$^{-2}$), respectively. The experimentally reported value of electrical conductivity for V at room temperature are 3.61$\times 10^{6}$ $\Omega^{-1}$m$^{-1}$\cite{pan1977investigation}, respectively. Therefore, according to Eq. \ref{eq.WFL}, $\kappa_{\rm{el}}$ of V at 300 K should be 26.42. Furthermore, the experimentally reported total thermal conductivity of V is 35.0 W/m-K at 300 K\cite{Jung}. Hence, lattice thermal conductivity ($\kappa_{\rm{ph}}$) at 300 K should be 8.57 W/m-K, which is in a good match with our calculated $\kappa_{\rm{ph}}$. Like 300 K, $\kappa_{\rm{ph}}$ is also extracted at different values of $T$ using the corresponding experimental values of $\sigma$\cite{pan1977investigation} and $\kappa$\cite{Jung}. The extracted values of $\kappa_{\rm{ph}}$ is shown in figure. Apart from the slight deviations, it is seen from the figure that the calculated values of $\kappa_{\rm{ph}}$ are very close to the extracted $\kappa_{\rm{ph}}$. 

Furthermore, lattice thermal conductivity of Nb is also calculated within the temperature range 100-1000 K. The corresponding results are depicted in Fig. \ref{Fig.kappa_Nb} along with the extracted $\kappa_{\rm{ph}}$. It is seen from the figure that the values of $\kappa_{\rm{ph}}$ decrease with the rise in $T$. At 300 K, the calculated value of $\kappa_{\rm{ph}}$ is 9.95 W/m-K, while it reaches 3.04 W/m-K at 1000 K. The experimentally reported value of $\sigma$ at room temperature is 5.91$\times 10^{6}$ $\Omega^{-1}$m$^{-1}$\cite{TYE196113}. Therefore, $\kappa_{\rm{el}}$ at 300 K should be 43.31 W/m-K, respectively. Moreover, the experimentally reported total thermal conductivity is 52.12 W/m-K at 300 K\cite{PhysRevB.28.6316}. Hence, lattice thermal conductivity ($\kappa_{\rm{ph}}$) Nb at 300 K should be 8.81 W/m-K, which is in a good agreement with our calculated $\kappa_{\rm{ph}}$. Similar to 300 K, $\kappa_{\rm{ph}}$ is also extracted at higher values of $T$ using the corresponding experimental values of $\sigma$\cite{moore1980thermal} and $\kappa$\cite{moore1980thermal}. The extracted values of $\kappa_{\rm{ph}}$ is shown in figure. From the figure, it is found that calculated $\kappa_{\rm{ph}}$ is in good match with the extracted $\kappa_{\rm{ph}}$ at the corresponding values of $T$.

 \begin{figure}\label{Fig.kappa_Nb}
\includegraphics[width=0.90\linewidth, height=5.0cm]{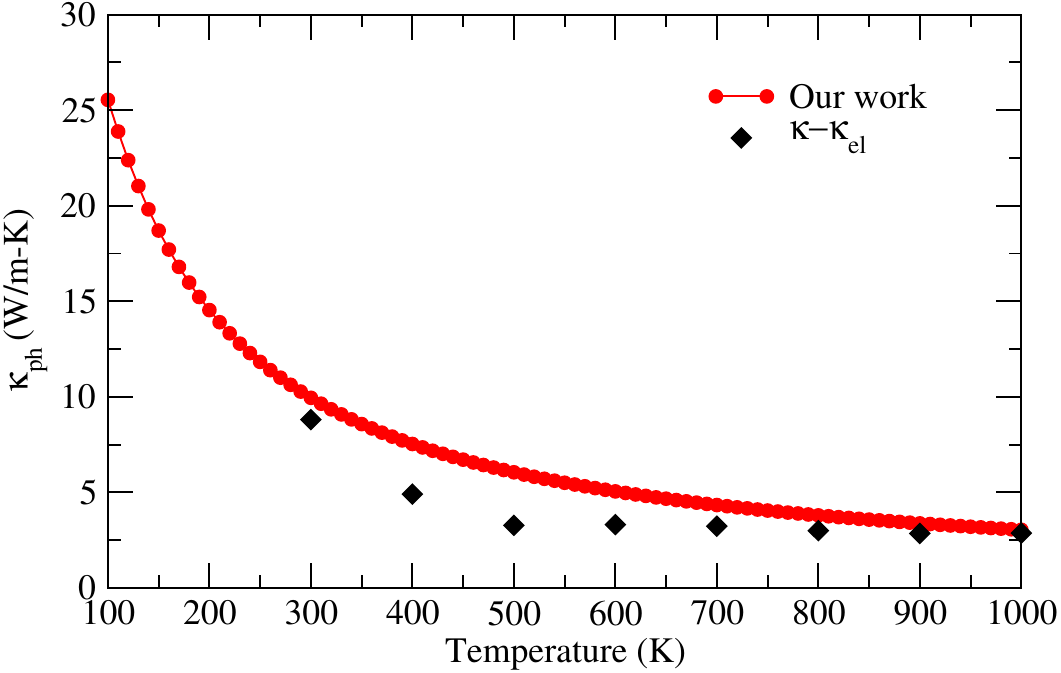}
\caption{\label{Fig.kappa_Nb}\small{Lattice thermal conductivity ($\kappa_{\rm{ph}}$) of Nb as a function of temperature. The splitted data points represent the extracted $\kappa_{\rm{ph}}$, which is obtained by subtracting the electronic contribution ($\kappa_{\rm{el}}$) from experimental thermal conductivity ($\kappa$) data.}}
\end{figure}

%***************************************** Conclusions **************************************
\section{Conclusions}
In summary, the phonon dispersion of BCC V and Nb is calculated using DFT-based methods and compared with the available experiments. The calculated results show a \enquote{dip}-like feature in the LA phonon mode along the $\Gamma$-H high symmetric path for both transition metals in supercell size 4$\times$4$\times$4. However, in supercell size 2$\times$2$\times$2 and 3$\times$3$\times$3, the \enquote{dip}-like feature is not clearly visible. This study clearly reveals that long-range interactions play a crucial role in phonon dispersion and related properties. The phononic part of Helmholtz free energy, specific heat at constant volume ($C_v^{\rm{ph}}$) and entropy ($S^{\rm{ph}}$) of these transition metals are calculated as a function of temperature under harmonic approximation. The calculated $C_v^{\rm{ph}}$ and $S^{\rm{ph}}$ are in good agreement with the available experimental results. Nextly, the phonon lifetime due to electron-phonon interactions ($\tau_{\rm{eph}}^{\rm{ph}}$) is calculated. The value of $\tau_{\rm{eph}}^{\rm{ph}}$ of V (Nb) is found to be $23.16\times 10^{-15}$ ($24.70\times 10^{-15}$) seconds (s) at $100$ K, which gets decreased to $1.51\times 10^{-15}$ ($1.85\times 10^{-15}$) s at 1000 K. In addition to this, the temperature-dependent phonon lifetime due to phonon-phonon interactions (PPI) are also calculated using the anharmonic approximation. The total phonon lifetime of V (Nb) due to PPI is found to be $8.59\times 10^{-12}$ ($18.09\times 10^{-12}$) and $0.83\times 10^{-12}$ ($1.76\times 10^{-12}$) s at 100 and 1000 K, respectively. Furthermore, the lattice thermal conductivity ($\kappa_{\rm{ph}}$) is computed using linearized phonon Boltzmann equation. At 300 K, the value of $\kappa_{\rm{ph}}$ is obtained for V (Nb) 9.93 (9.95) W/m-K. It is found that calculated $\kappa_{\rm{ph}}$ is in good match with the extracted $\kappa_{\rm{ph}}$ at the corresponding values of $T$. This study states that variation of phonon dispersion with supercell size is crucial for understanding the phonon properties of solids accurately.

\bibliography{MS}
\bibliographystyle{apsrev4-2}

\end{document}